\documentclass[12pt]{article}
\usepackage{epsfig}

\usepackage{amssymb}
\usepackage{amsfonts}

\def\be{\begin{equation}}
\def\ee{\end{equation}}
\def\ba{\begin{array}{c}}
\def\ea{\end{array}}

\def\ben{$$}
\def\een{$$}
\newcommand{\bbr}{\br\!\br}

\newcommand{\kt}{\rangle}
\newcommand{\br}{\langle}
\begin{document}


\vspace{.35cm}

 \begin{center}{\Large \bf

Three solvable matrix models

of a quantum catastrophe

 }

  \end{center}

\vspace{10mm}

 \begin{center}

 {\bf G\'eza L\'evai}

 \vspace{3mm}

ATOMKI,

Debrecen, Hungary

{e-mail: levai@atomki.mta.hu}

 \vspace{3mm}

 and

  {\bf Franti\v{s}ek R$\!$\accent23{$\!\!\!\!$u}\v{z}i\v{c}ka} and
 {\bf Miloslav Znojil}

 \vspace{3mm}
Nuclear Physics Institute ASCR,

250 68 \v{R}e\v{z}, Czech Republic

{e-mail: znojil@ujf.cas.cz}

\vspace{3mm}


\end{center}

\vspace{5mm}

\newpage

\section*{Abstract}

Three classes of finite-dimensional models of quantum systems
exhibiting spectral degeneracies called quantum catastrophes are
described in detail. Computer-assisted symbolic manipulation
techniques are shown unexpectedly efficient for the purpose.

\subsection*{Keywords}

quantum theory; PT symmetry; finite-dimensional non-Hermitian
Hamiltonians; exceptional-point localization; quantum theory of
catastrophes; methods of computer algebra;

\newpage

\section{Introduction}


For non-specialists, the presentation of some of the key ideas of
quantum theory may be mediated by the technically next-to-trivial
toy-model quantum systems ${\cal S}_{(N=2)}$ in which the
representation of a dynamical state, i.e., in the Dirac's notation,
of the ket-vector element $|\psi^{(P)}\kt$ of a pre-selected
physical Hilbert space ${\cal H}^{(P)}$ is just a two-component
complex vector
 \be
 |\psi^{(P)}_{(N=2)}\kt=
 \left (
 \ba
  \psi^{(P)}_1\\
  \psi^{(P)}_2
  \ea
  \right )\,
  \in \,
  {\cal H}^{(P)}_{(N=2)}\,,
  \ \ \ \ \
  {\cal H}^{(P)}_{(N=2)}\ \equiv \
  \mathbb{C}^2
   \,.
   \label{twodime}
  \ee
Exaggerated as such a drastic reduction might seem, it still may be
found in various analyses of the meaning and modern interpretations
of quantum theory. In particular, precisely this simplification has
been chosen by Dorey et al \cite{DDT} as playing a crucial
methodical role in the running development of the so called ${\cal
PT}$-symmetric quantum mechanics (PTQM) in 2001.

One should remember that the ultimate formulation of PTQM theory was
only completed a few years later (cf. \cite{Carl,ali} and also
\cite{SIGMA} for more details). The key conceptual problems
encountered during the introduction and development of the PTQM
physics appeared difficult, albeit more or less purely technical.
The work with $N-$dimensional analogues of  Eq.~(\ref{twodime}) at
suitable finite $N < \infty$ proved helpful.

Some of the problems emerged immediately after Bender and Boettcher
\cite{BB} proposed, in 1998, a generalization of the usual
description of the unitary evolution based on the  current physical
Schr\"{o}dinger equation
 \be
 {\rm i}\partial_t\,|\psi^{(P)}\kt=\mathfrak{h}\,|\psi^{(P)}\kt\,
  \label{SEti}
 \ee
containing a Hermitian Hamiltonian $\mathfrak{h}
=\mathfrak{h}^\dagger$. They declared that the explicit postulate of
the Hermiticity of the Hamiltonian had to be weakened. Such an
attempt opened the Pandora's box of complicated mathematical
questions, some of which remain unresolved up to these days (cf.,
e.g., \cite{Siegl}).
%
%

On positive side, the PTQM framework was found mathematically
correct (it is briefly summarized in Appendix A below). {Under
suitable formal restrictions } the formulation of which dates back
to paper \cite{Geyer} by Scholtz et al in 1992, the potential
applications of the theory do not even require any excessively
complicated mathematics.

We intend to demonstrate that the PTQM formalism is exceptionally
suitable for a computer-assisted model building. Having chosen a
triplet of suitable $N$ by $N$ matrix examples (cf. their
introduction in section \ref{inifta}) we shall present their
detailed methodical analysis in a fully constructive and
non-numerical, symbolic-manipulation-based spirit.

The key purpose of our message will lie in the demonstration of the
not quite obvious feasibility of constructions at all dimensions. We
shall assume  that our Hamiltonians $H \neq H^\dagger$ can vary with
one or more free parameters. This will open perspectives in analysis
of less common quantum systems ${\cal S}$.

We shall consider the three series of $N$ by $N$ matrix models as
initiated by the respective $N=2$ Hamiltonian matrices
 \be
  \left [\begin {array}{cc}
 2-{\rm i}\lambda&-1\\{}-1\ &2+{\rm i}\lambda
 \end {array}\right ]
 \,,\ \ \
  \left [\begin {array}{cc}
 2&-1+\lambda\\{}-1-\lambda&\ 2
 \end {array}\right ]
 \,,\ \ \
   \left [\begin {array}{cc}
 2&-1+\lambda\\{}-1-\lambda&\ 2
 \end {array}\right ]
 \,
 \label{list3}
 \ee
(notice that the second and the third items coincide at $N=2$). In
all of these  $N=2$ initial cases the determination of the bound
state energies remains trivial yielding, incidentally, the same
result,
 \ben
 E_\pm=2\pm \sqrt{1-\lambda^2} \,.
 \een
Surprisingly enough, we shall reveal that for all of our toy models,
the facilitated tractability of spectra survives the generalizations
$(N=2) \to (N>2)$. At any matrix dimension our examples will share
the property of possessing a compact domain of acceptable physical
parameters $\lambda \in {\cal D}$ at which the spectrum is real and
non-degenerate. The $N>2$ descendants $H^{(N)}$ of the above three
$N=2$ matrices will also share a number of further qualitative
features. One of the important ones is that our Hamiltonians will
cease to be diagonalizable at the boundary points of the respective
domains ${\cal D}$. According to Kato \cite{Kato}, these boundary
points  may be called exceptional points (EP). We may also see that
$ {\cal D}= (-1,1)$ for $N=2$.

From the point of physics every EP manifold $\partial {\cal
D}^{(N)}$ forms, in the space of parameters, a natural horizon of
observability of the system. These horizons may intuitively be
perceived as formal singularities {\em alias } instants of the
physical system's collapse. Depending on a more concrete
interpretation of the system they may also carry the meaning of
phase transition or of a  cusp-like quantum catastrophe (QC,
\cite{I}). In section \ref{fourthmo} we shall briefly return to this
point and to the results of Ref.~\cite{I} which will be cited as
paper I in what follows. These results will serve us as a methodical
guide because, in particular, the underlying $N=2$ Hamiltonian
 \ben
  \left [\begin {array}{cc}
 1&\ \lambda\\{}-\lambda&\ 3\end {array}\right ]\,
 \een
might have very easily been added to the above list (\ref{list3}) as
a fourth item.

In the rest of the paper we shall pay detailed attention to the new
models and to the role of the computer-assisted symbolic
manipulations in their analysis. In particular, in section
\ref{infinita} we shall sample the secular equations of the three
new classes of $N$ by $N$ matrix models. As long as we will not
impose restrictions on the size of dimension $N$, we shall argue
that the computerized manipulations may be expected vital for the
feasibility and success of the search for the parameters and spectra
in the vicinity of the QC singularities.

In section \ref{reinifta} we shall re-attract the reader's attention
to an interplay between the topologically anomalous EP degeneracies
of the spectra and certain geometrical features of the related
physical Hilbert space. We shall explain how the computer-assisted
manipulations can mediate or improve our insight into the
mathematical structures of quantum systems which are close to the QC
dynamical regime.

In the last section \ref{discussion}
a few comments on the purpose of our  toy-model simulations and
on the possible physics
of
quantum systems near QC will be added.

\section{The QC menu of $N$ by $N$ matrix models\label{inifta}}

In paper I one of us studied the cusp-like quantum catastrophes at
$\lambda \in \partial { \cal D}^{(N)}$ via an analysis of specific
toy models. In respective paragraphs \ref{firstm} - \ref{thirdm} let
us now introduce three other illustrative QC-generating models.

\subsection{The first series of models:
Systematic discrete approximants of differential-operator
Hamiltonians\label{firstm}}

People often restrict the class of interesting PTQM Hamiltonians to
the most common superposition $H=T+V$ of a kinetic-energy Laplacean
$T\sim -\triangle$ with a local potential $V=V(x)$ \cite{BB}. For
analogous Hamiltonians $H^{(N)}$ given in a finite-dimensional
matrix form, it is most natural to insert, in the same formula
$H^{(N)}=T^{(N)}+V^{(N)}$, the free-motion form of the discrete
Laplacean 
 \be
  T^{(N)}=  \left[ \begin {array}{ccccc}
2&-1&0&\ldots&0
\\
-1&\ \ \ 2\ \ \ &\ddots&\ddots&\vdots
\\
0&\ddots&\ \ \ \ddots\ \ \ &-1&0
\\
\vdots&\ddots&-1&2&- 1
\\
0&\ldots&0&-1&2
\end {array}
 \right]\,.
 \label{toymkin}
 \ee
%
%
This matrix must be complemented by a suitable real and confining
local interaction defined along an underlying discrete equidistant
coordinate lattice $x_j \sim hj +const$. Here, $h>0$ is a given
grid-point distance. The construction leads to the family of
grid-point models (GPM)
 \be
 {H}^{(N)}=
 \left (
 \begin{array}{ccccc}
 2+h^2V(x_{1})&-1&&&\\
 -1&2+h^2V(x_{2})&-1&&\\
  &-1\ \ \ & 2+h^2V(x_{3}) & \ddots&\\
  & & \ddots\ \ \ \ \ &\ddots&-1 \\
 &&&-1\ \ \ \ &2+h^2V(x_{N})
 \ea
 \right )\,.
 \label{dvanact}
 \ee
Once we restrict our attention to the discrete analogues of the
purely imaginary cubic oscillator potential $V(x)={\rm i}x^3$ of
Ref.~\cite{BB}, our GPM Hamiltonians become ${\cal PT}-$symmetric
(cf. Appendix A), with the parity-transformation-representing real
matrix
 \be
 {\cal P}= {\cal P}^{(N)}=
  \left[ \begin {array}{ccccc}
 0&0&\ldots&0&1
 \\{}0&\ldots&0&1&0\\
 {}\vdots&
 {\large \bf _. } \cdot {\large \bf ^{^.}}&
 {\large \bf _. } \cdot {\large \bf ^{^.}}
 &
 {\large \bf _. } \cdot {\large \bf ^{^.}}&\vdots
 \\{}0&1&0&\ldots&0
 \\{}1&0&\ldots&0&0
 \end {array} \right]\,.
 \label{anago}
 \ee
The necessarily antilinear operator ${\cal T}$ may be prescribed by
formula (\ref{Dirac}) of Appendix A. For illustration, a number of
semi-qualitative spectrum-analyzing results of GPM systems may be
found, e.g., in Ref.~\cite{annals}.

\subsection{The second model: Non-local  interaction
at the boundary\label{bounda}}

A serious conceptual weakness of GPM Hamiltonians (\ref{dvanact})
emerges when one tries to move to the scattering dynamical regime
(cf. \cite{Jones}). In Ref.~\cite{scatt} a consistent conceptual
remedy of this weakness has been found in a transition to suitable
non-local interactions $V(x,x')$. The transition to non-local
interactions converts realistic Hamiltonians $H=T+V$ into
integro-differential operators at $N=\infty$. This forces us to pick
up the exceptionally simple potentials $V$ of course.

One of the best and simplest differential-operator candidates is
provided by the boundary-interaction model (BIM) of
Ref.~\cite{david}. A discrete, approximative $N<\infty$ version of
such a BIM choice of $H$ has been analyzed in Ref.~\cite{davidium}.
We argued there that for the sake of simplicity one should prefer
just minimally non-diagonal, tridiagonal matrices $V(x_j,x'_k)$ for
illustration purposes. We decided to select such a special
Schr\"{o}dinger equation  $H^{(N)}(\lambda)|\psi\kt =E\,|\psi\kt$
which is best presented in the following tri-partitioned matrix form
 \be
   \left[ \begin {array}{cc|cc|cc}
 2-E&-1-{\it {\lambda}}&0&\ldots&0&0
\\
{}-1+{\it {\lambda}}&2-E&-1&0&\ldots&0
\\
\hline
 {}0&-1&\ \ \ 2-E\ \ \ &\ddots&\ddots&\vdots
\\
{}\vdots&0&\ddots&\ \ \ \ddots\ \ \ &-1&0
\\
\hline {}0&\vdots&\ddots&-1&2-E&- 1+{\it {\lambda}}
\\
{}0&0&\ldots&0&-1-{\it {\lambda}}&2-E
\end {array}
 \right]\,
\left[ \begin {array}{c}
 \psi_1\\
 \psi_2\\
 \hline
 \psi_3\\
 \vdots\\
 \psi_{N-2}\\
 \hline
 \psi_{N-1}\\
 \psi_{N}
 \end {array}
 \right]=0\,.
 \label{toymSE}
 \ee
As long as the interaction is non-local, a few further comments
on its properties are due and may be found in Appendix B below.

\subsection{The third option: Multi-parametric discrete square
wells\label{thirdm}}

In paper \cite{Junde} the authors felt inspired by the formal
parallels between local and non-local interactions supported at the
endpoints of the spatial lattice. One of us further generalized the
interaction matrix $V$ which was now left non-vanishing far from the
two ends of the lattice \cite{bJunde}. In the continuous coordinate
limit $N \to \infty$ this may still simulate certain less elementary
boundary conditions but the overall perception of the interaction is
different, more general. At the finite dimensions $N<\infty$, the
interaction Hamiltonians of the latter type may be called here
nearest-neighbor-interaction models (NNIM). In their spirit we may
now replace Eq.~(\ref{toymSE}) by its ``maximal'', two-parametric
generalization at $N=4$,
 \be
 H^{(4)}(\vec{\lambda})=  \left[ \begin {array}{cccc}
 2&-1+\beta&0&0\\\noalign{\medskip}-1-
 \beta&2&-1+\alpha&0\\\noalign{\medskip}0&-1-\alpha&2&-1+\beta
 \\\noalign{\medskip}0&0&-1-\beta&2\end {array} \right]\,
 \label{hamal}
 \ee
etc. In general one may demand either
 \begin{eqnarray}
 V_{21}=-V_{12}=V_{N-1,N}=-V_{N,N-1}
 =\lambda\,, \nonumber \\
 V_{23}=-V_{32}=V_{N-1,N-2}=-V_{N-2,N-1}
 =\mu\,, \\
 V_{43}=-V_{34}=V_{N-3,N-2}=-V_{N-2,N-3}
 =\nu\,, \nonumber \\
 \ldots\,.\nonumber
 \label{tripar}
 \end{eqnarray}
or
 \begin{eqnarray}
 V_{21}=-V_{12}=-V_{N-1,N}=
 V_{N,N-1}
 =\lambda\,, \nonumber \\
 V_{23}=-V_{32}=-V_{N-1,N-2}=V_{N-2,N-1}
 =\mu\,, \\
 V_{43}=-V_{34}=-V_{N-3,N-2}=V_{N-2,N-3}
 =\nu\,, \nonumber \\
 \ldots\,\nonumber
 \label{triparbez}
 \end{eqnarray}
Without any real change of the model's phenomenological appeal, the
surviving correspondence with the $N=\infty$ differential-operator
partners is more flexible. In spite of that, the closed-form
solvability need not be lost.

\section{Non-Hermiticity of  Hamiltonians revisited\label{fourthmo}}

\subsection{Elementary toy model: Discrete ${\cal PT}-$symmetric
anharmonic oscillator \label{fourthm}}

In paper \cite{maximal} one of us revealed that the family of
toy-model Hamiltonians
 \be
 H^{(N)} =
 \left [\begin {array}{ccccc}
 \ 1\  &0  &0  &\ldots&0\\
 0&\ 3 \  &0  &\ldots&0 \\
  0&0  &5 &\ddots& \vdots\\
  \vdots&\vdots&\ddots&\ddots&0\\
  0&0&\ldots&0&2\!N\!\!-\!\!1
 \end {array}\right ]
 +g\,V^{(N)}
 \label{akvence}
  \ee
is most naturally interpreted, at the sufficiently small couplings
$g$, as a diagonalized and truncated harmonic oscillator which is
complemented by a suitable perturbation $g\,V^{(N)}$. The
perturbation was assumed maximally elementary {\em and} maximally
non-Hermitian. Tridiagonal and real matrices $V^{(N)}$ were chosen,
therefore, antisymmetric and pseudo-Hermitian with respect to the
parity-simulating $N$ by $N$ matrix operator (\ref{anago}).

The most interesting one-parametric version of this
anharmonic-oscillator model (AOM) has been found represented by
$N-$numbered interaction matrices
 \ben
 V^{(2)}_{} = \left [\begin {array}{cc}
 0&1\\{}-1&0\end {array}\right ]
 \,,\ \ \ \
 V^{(3)}_{} = \left [\begin {array}{ccc}
 0&\sqrt{2}&0\\
 {}-\sqrt{2}&0&\sqrt{2}\\
 0&-\sqrt{2}&0
 \end {array}\right ]
 \,,\ \ \ \
 \een
 \be
 V^{(4)}_{} = \left [\begin {array}{cccc}
  0&\sqrt{3}   &0  &0\\
 -\sqrt{3} &0   &2  &0\\
  0&-2  &0 &\sqrt{3} \\
  0&0&-\sqrt{3} &0
 \end {array}\right ]\,,\ \ldots\,.
 \label{sequence}
 \ee
From our present QC-oriented point of view the most important
property of such a family of models is that in the strong-coupling
regime, they all admit a remarkable reparametrization of coupling $g
= \sqrt{1-t} $, yielding  equidistant spectrum which ``blows up''
with $t\geq 0$,
 \be
 E_n(t)=(2n+1)\sqrt{t}+ const\,,\ \ \ \ n = 0, 1, \ldots, N-1
 \,.
 \ee
At any  $N < \infty$ it is real for  positive ``time''  $t>0$ while
it totally degenerates to the single real value $E_n(0)=const$ at
$t=0$. At negative $t<0$ this spectrum loses {\em both} the
degeneracy and the observability (i.e., reality).

\subsection{A rederivation of the maximal QC
confluence\label{fourthmeoe}}

In the spirit of paper I the latter observations characterize a QC
regime. These results are to be extended now to the other
Hamiltonians. Yet before doing so, let us still return, briefly, to
the algorithms for the derivation of the QC-producing interactions
(\ref{sequence}), restricting attention to the  even dimensions
$N=2K$. In comparison with the study as performed in paper I, we
intend to shift emphasis from the qualitative needs of quantum
phenomenology to its constructive and quantitative aspects and, in
particular, to the merits of applicability of the methods of
computational linear algebra. In this spirit, let us now review some
vital technicalities.

The first step of a re-derivation of Eq.~(\ref{sequence}) is a
constant shift of energies plus their square-root re-parametrization
$E-const = \pm \sqrt{s}$ (i.e., a symmetrization of the spectrum
with respect to the origin). Under the choice of NNIM Hamiltonian
with $K=1$ this yields the simplified secular equation
 \be
 \det \left (H^{(2)}-(E-2)\right )=
 \det \left[ \begin {array}{cc} -1-\sqrt {s}&\sqrt {\alpha}\\
 \noalign{\medskip}-\sqrt {\alpha}&1-\sqrt {s}\end {array} \right]=0\,
 \ee
which proves reducible to the utterly elementary explicit formula
$s=1-\alpha$. The nontrivial challenge only comes at the next
dimension $N=2K=4$ with \be
 \det \left (H^{(4)}-(E-4)\right )=
 \det \left[ \begin {array}{cccc} -3-\sqrt {s}&\sqrt {\beta}&0&0\\
 \noalign{\medskip}-\sqrt {\beta}&-1-\sqrt {s}&\sqrt {\alpha}&0
\\\noalign{\medskip}0&-\sqrt {\alpha}&1-\sqrt {s}&\sqrt {\beta}
\\\noalign{\medskip}0&0&-\sqrt {\beta}&3-\sqrt {s}\end {array}
 \right]
 =0\,.
 \ee
This secular equation is reducible to the
exactly solvable quadratic equation
 $${s}^{2}+ \left( -10+2\,\beta+\alpha \right)
 s+9+6\,\beta-9\,\alpha+{
 \beta}^{2}
 =0.$$
Without even solving it explicitly, this equation proves suitable
for the tests of applicability and efficiency of computer-assisted
symbolic manipulations.

In a methodically motivated search for domain ${\cal D}$ we would
have to start from the re-derivation of the QC-related parameters as
given in Eq.~(\ref{sequence}). Before knowing the ultimate,
$N-$times confluent QC singularity in explicit form we can only say
that all of the QC energy roots of the secular equation must
degenerate to zero in such a limit ($s=0$). This means that at any
$K$ one has to satisfy a multiplet of polynomial equations. They
should be solved by the Gr\"{o}bner-basis techniques.

The procedure is lengthy and may be found described in
\cite{maximal}. In its $N=4$ illustration the coupled set is merely
formed by two polynomial equations
 \be
 -10+2\,\beta+\alpha=0\,,\ \ \ \ 9+6\,\beta-9\,\alpha+{
\beta}^{2}=0\,. \ee As long as $N=2K=4$, one does not need any
software to solve the problem. Incidentally, the solution may be
also obtained in closed form and proved unique at all $N$. For
another NNIM explicit illustration of algebra let us now put  $K=3$
and get the secular equation
 \ben
 {s}^{3}+ \left( -35+2\,\gamma+\alpha+2\,\beta \right) {s}^{2}+
 \een
 \ben
 +\left(
 -34\,\alpha+2\,\alpha\,\gamma+259+{\beta}^{2}
 +{\gamma}^{2}+28\,\gamma+
 2\,\beta\,\gamma-44\,\beta \right) s-
 \een
 \ben
 -225-30\,\gamma+30\,\alpha\,\gamma
 -10\,\beta\,\gamma-150\,\beta-25\,
 {\beta}^{2}+\alpha\,{\gamma}^{2}+225
 \,\alpha-{\gamma}^{2}=0\,.
 \een
By its inspection, the unavoidable necessity of symbolic
manipulations at $N \geq 6$ becomes entirely obvious.

\section{Secular equations in closed form\label{infinita}}

\subsection{The GPM models of paragraph \ref{firstm}\label{firstmo}}

For the first nontrivial $N=3$ GPM Hamiltonian of
Ref.~\cite{annals},
 $$
 H^{(3)}=H^{(3)}(\alpha)=\left[ \begin {array}{ccc} -i\alpha&-1&0\\
 \noalign{\medskip}-1&0&-1\\
 \noalign{\medskip}0&-1&i\alpha\end {array} \right]
 $$
the triplet of energies is easily obtained in closed form, with
$E_0=0$ while $E_\pm = \pm \sqrt {2-{\alpha}^{2}}$. Alas, such a
model is merely one-parametric. With interval ${\cal
D}^{(2)}=(-\sqrt{2},\sqrt{2})$ and with just the two isolated
boundary EPs, it is still too elementary.

At the next dimension $N=4$ let us contemplate GPM Hamiltonian
 \be
 H^{(4)}=\left[ \begin {array}{cccc} -i\alpha&-1&0&0\\
 \noalign{\medskip}-1&-i\beta&-1&0\\
 \noalign{\medskip}0&-1&i\beta&-1\\\noalign{\medskip}0&0&-1
&i\alpha\end {array} \right]\,
\label{difere}
 \ee
and let us simplify $E=\sqrt{s}$. The resulting secular equation
 \be
 {{\it s}}^{2}+ \left( {\alpha}^{2}+{\beta}^{2}-3 \right) {{\it s}}^{}
 +{\alpha}^{2}{\beta}^{2}+2\,\alpha\,\beta-{\alpha}^{2}+1
 =0\,
 \label{sec4z}
 \ee
yields the energy eigenvalues in closed form. Still, its
two-parametric nature already provides a nontrivial methodical
guidance for a move to any matrix dimension $N$. It may be
summarized as follows. First of all, while starting from matrix
dimension $N=4$ it makes sense to pre-assist the algebraic
manipulations by recalling some graphical software. A rough
orientation can be obtained concerning both the topology of spectra
{\em and} of the EP boundaries $\partial {\cal D}$. A characteristic
sample of the latter structure is provided by
Fig.~\ref{picee3ww}. 

\begin{figure}[h]                     
\begin{center}                         
\epsfig{file=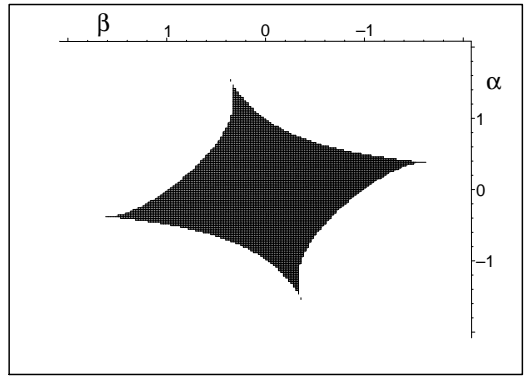,angle=270,width=0.5\textwidth}
\end{center}                         
\vspace{-2mm}\caption{In the $\alpha-\beta$ plane the eigenvalues of
matrix (\ref{difere}) only remain real inside the black domain.
 \label{picee3ww}}
\end{figure}

Starting from $N=4$ it makes also sense to search for the multiply
and, in particular, maximally degenerate EPs (MEPs), i.e., for the
vertices of the spikes as seen in our two-dimensional
Fig.~\ref{picee3ww}. The required algorithm may be entirely
universal. Indeed, as long as all of the roots $s_n$,
$n=1,2,\ldots,N$  of our secular equation must degenerate to zero at
MEPs, each coefficient in the secular polynomial must vanish (up to
the leading one of course). One obtains a multiplet of polynomial
equations, perfectly fitting the capabilities of the
Gr\"{o}bner-basis reduction and elimination techniques \cite{MAPLE}.

The $N=4$ example is already sufficiently instructive because its
two MEP conditions derived from secular Eq.~(\ref{sec4z}) form a
nontrivial coupled pair of polynomial equations
 \be
 \alpha^2+\beta^2=3\,,\ \ \ \ \ \ \ \
 \alpha^2=(1+\alpha\beta)^2\,.
 \ee
Under a proper account of symmetries the Gr\"{o}bner-basis
elimination provides the effective quartic polynomial equation
 $$
 2-6\,\beta+2\,{\beta}^{2}+2\,{\beta}^{3}-{\beta}^{4}=0
 $$
possessing the two complex and two real MEP roots. Their numerical
localization enables us, finally, to find the exact coordinates of
the four spikes in Fig.~\ref{picee3ww}, viz.,
$$\beta= \pm 1.691739510\,,\ \ \ \ \ \ \alpha=\mp 0.3715069717$$
$$\beta=\pm 0.4060952085\,,\ \ \ \ \ \
\alpha=\pm 1.683771565.$$
%
%
No genuine methodical news appear during the transition to the more
realistic as well as more complicated secular equations at any
larger $N>4$.

\subsection{The new BIM and NNIM models of paragraphs \ref{bounda}
and \ref{thirdm}, respectively\label{boundao} }

As long as the one- or few-parametric GPM Hamiltonian matrices of
paragraph \ref{bounda} may be perceived as mere special cases of the
fully general multi-parametric NNIM $N$ by $N$ matrices of paragraph
\ref{thirdm}, both of these dynamical scenarios may be studied in
parallel. Also the majority of observations of subsection
\ref{firstmo} may be taken over without truly essential changes.
Thus, for our first nontrivial $N=4$ illustrative example
(\ref{hamal}) the energy spectrum coincides with the roots of the
following elementary secular equation
 $$
 {{\it E}}^{4}+ \left( {\alpha}^{2}-3+2\,{\beta}^{2} \right)
  {{\it E}}^{2}+1-2\,{\beta}^{2}+{\beta}^{4} =0\,.
 $$
These energies will occur in pairs $E_{\pm,\pm}=\pm \sqrt{Z_\pm}$
where the symbol $Z_\pm$ denotes the two easily deduced roots of
quadratic equation. As above, the latter two roots must be
non-negative inside the closure of the physical parametric domain
${\cal D}$. Thus, {\em mutatis mutandis} one can repeat the
construction steps as made in preceding subsection. At any dimension
$N$ these steps involve again the comprehensive graphical analysis
of the energy spectra as well as the explicit Gr\"{o}bner-basis
constructions of the MEP parameters.

Our enthusiasm over the efficiency of algorithms gets enhanced by
the wealth of complexity of the topological structure of the
spectral loci (cf. their graphical samples in Ref.~\cite{Junde}).
Similar pleasant surprises also emerge in the MEP-construction
perspective. Let us  display one of the results in which the tedious
Gr\"{o}bner-basis constructive effort resulted in an amazing $N \to
N+1$ extrapolation hypothesis. The essence of this new scenario is
best illustrated by the following $N=10$ example
 \be
  \left[ \begin {array}{cccccccccc} 2&-1-t&0&0&0&0&0&0&0&0
\\\noalign{\medskip}-1+t&2&-1+t&0&0&0&0&0&0&0\\\noalign{\medskip}0&-1-
t&2&-1-t&0&0&0&0&0&0\\\noalign{\medskip}0&0&-1+t&2&-1+t&0&0&0&0&0
\\\noalign{\medskip}0&0&0&-1-t&2&-1-t&0&0&0&0\\\noalign{\medskip}0&0&0
&0&-1+t&2&-1+t&0&0&0\\\noalign{\medskip}0&0&0&0&0&-1-t&2&-1-t&0&0
\\\noalign{\medskip}0&0&0&0&0&0&-1+t&2&-1+t&0\\\noalign{\medskip}0&0&0
&0&0&0&0&-1-t&2&-1-t\\\noalign{\medskip}0&0&0&0&0&0&0&0&-1+t&2
\end {array} \right]\,
 \label{zaklad}
 \ee
for which the  $t-$dependence of the spectrum is shown in
Fig.~\ref{fi1}.

\begin{figure}[h]                     
\begin{center}                         
\epsfig{file=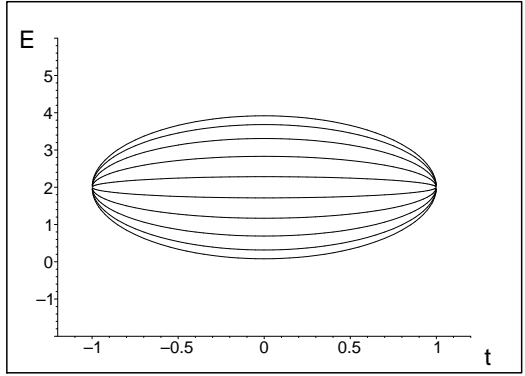,angle=270,width=0.5\textwidth}
\end{center}                         
\vspace{-2mm}\caption{Real eigenvalues of Hamiltonian (\ref{zaklad})
exhibiting the two phase-transition MEP  complexification singularities at
$t=\pm 1$.
 \label{fi1}}
\end{figure}

\section{Remarks on the constructions of metrics\label{reinifta}}


The explicit constructive assignment of a nontrivial metric
$\Theta\neq I$ to a given Hamiltonian $H$ with real spectrum is not
easy even at $N < \infty$ and even with the assistance of computers.
In one of the most ambitious approaches to this problem (as accepted
also in paper I) the general assignment of a metric $\Theta$ to the
given Hamiltonian $H$ relies on the brute-force solution of the set
of $N^2$ linear-algebraic equations (\ref{dieudo}) of Appendix A.
The technical details of such an assignment remain nontrivial (cf.
\cite{CASC}).

There exist several successful strategies of the  explicit
evaluation of solutions $\Theta=\Theta(H)$. As mentioned, one of
them consists in a linear-algebraic approach to Eq.~(\ref{dieudo})
of Appendix A. It may prove to be the most efficient one, especially
in sparse-matrix models \cite{recurrent}. An alternative recipe is
offered by the  spectral-type expansion formula
 \be
 \Theta=\Theta(H,\vec{\kappa})=\sum_{j=0}^{N-1}\,
 |\Xi_j\kt\,\kappa_n\,\br \Xi_j|\,
 \label{metrr}
 \ee
where symbols $|\Xi_j\kt$ denote the eigenkets of the conjugate operator
$H^\dagger$ while the optional parameters $\kappa_j>0$ remain
arbitrary \cite{SIGMAdva}.

For illustration let us now return, once more, to the discrete AOM
quantum system. The model can only be declared completed after the
Hamiltonian $H$ had been complemented by the selection of the
Hilbert-space metric $\Theta$. In paper I we showed that the
construction of such a metric is, with the assistance of
sophisticated computer-mediated symbolic manipulations, feasible
even if complicated. The brute-force strategy with $N^2$ linear
equations has been recommended there. Due to the natural limitations
of the display of formulae we merely replaced the printout of
metrics by the  more compact (and, for reconstruction, sufficient)
presentation of the mere $N-$plets of eigenvectors $|\Xi_j\kt$.

Once we now have to move to a sample application of the two
alternative recipes to the three new models, let us pick up just the
one of paragraph \ref{thirdm} for the sake of definiteness. In
contrast to paper I where we were mainly interested in the {\em
physics} of quantum catastrophes (for which the knowledge of the
metric proved essential), we are now mainly interested in the  same
problem from the point of view of {\em computer-assisted
mathematics}. While our present main interest remained concentrated
on the genuinely {\em non-linear} problems of section
\ref{infinita}, such a change of point of view also implies a
definite weakening  of our interest in the {\em linear-algebraic}
problem of solution of Eq.~(\ref{dieudo}) for the metric.

This being said we may now recall our particular $N=4$ illustrative
example (\ref{hamal}) and reveal that the construction of the metric
leads to challenging problems. Let us explain the essence of one of
them which refers to both of the above alternative approaches. The
core of the message will lie in the construction which assigns the
two different metrics to the same Hamiltonian in dependence of the
choice of parameters from inside ${\cal D}$.

In the first step, in the spirit of Ref.~\cite{recurrent} one can
show that for any pair of ``small'' dynamical parameters $\alpha \in
(-1,1)$ and $\beta\in (-1,1)$ one of the metrics for Hamiltonian
(\ref{hamal}) may be chosen diagonal (i.e., $\Theta_{ij}=0$ iff $i
\neq j$). This choice is well defined and unique so that any other
non-equivalent positive definite metric $\Theta(H)$ must necessarily
acquire a non-diagonal matrix form.

Now one reveals a paradox by picking up a special version of quantum
Hamiltonian (\ref{hamal}) in which we choose a ``small''
$\alpha=\alpha_0=1/2$ and ``large'' $\beta=\beta_0=3\sqrt{2}/4
\approx 1.060660172 > 1$. Then we easily verify that the whole
spectrum of energies remains real and that it may even be written in
closed form, $E_{\pm,\pm}=(\pm 1\pm \sqrt{3})/4$. Still, as long as
$1.060660172>1$, the simplest possible diagonal-metric anssatz would
not work anymore in this regime -- it will simply fail to remain
positive definite. Such a conclusion declares the use of the
brute-force method useless in the whole domain of parameters ${\cal
D}$. In some of its parts, a return to the robust spectral-like
formula (\ref{metrr}) emerges as unavoidable.

As long as the eigenvectors of $H^\dagger$ as needed in
(\ref{metrr}) are not mutually orthogonal, the alternative metrics
need not be sparse matrices anymore. The resulting closed and
entirely general four-parametric form (\ref{metrr}) of metric
$\Theta=\Theta(H,\vec{\kappa})$ cannot be printed on a single page
as a consequence. Thus, one only has to display the explicit
eigenvectors in place of the whole matrix of metric. For
illustration we may pick up the sample values of
$\alpha=\alpha_0=1/2$ and $\beta=\beta_0=3\sqrt{2}/4$ and show that
the resulting eigenvectors still remain compact and easily
displayed,
 \be
 |\Xi_{++}\kt=\left (
 \ba
 2+3\,\sqrt {2}-2\,\sqrt {3}-6\, \left( 1/4+1/4\,\sqrt {3} \right)
 \sqrt {2}\\
 1\\
 -1/3\,\sqrt {3}\\
 2-2/3\,\sqrt {3}-2\,\sqrt {2}+2\, \left( 1/4+1/4\,\sqrt {3} \right)
 \sqrt {2}
 \ea
 \right )\,,
 \ee
%
%
%
%
%
%
%
%
 \be
 |\Xi_{+-}\kt=\left (
 \ba
 2+3\,\sqrt {2}+2\,\sqrt {3}-6\, \left( 1/4-1/4\,\sqrt {3} \right)
 \sqrt {2}\\
 1\\
 1/3\,\sqrt {3}\\
 2+2/3\,\sqrt {3}-2\,\sqrt {2}+2\, \left( 1/4-1/4\,\sqrt {3} \right)
 \sqrt {2}
 \ea
 \right )\,,
 \ee
%
%
%
%
%
%
 \be
 |\Xi_{-+}\kt=\left (
 \ba
 -2-3\,\sqrt {2}-2\,\sqrt {3}-6\, \left( -1/4+1/4\,\sqrt {3} \right)
 \sqrt {2}\\
 1\\
 -1/3\,\sqrt {3}\\
 2+2/3\,\sqrt {3}-2\,\sqrt {2}-2\, \left( -1/4+1/4\,\sqrt {3}
 \right) \sqrt {2}
 \ea
 \right )\,,
 \ee
%
%
%
%
%
%
 \be
 |\Xi_{--}\kt=\left (
 \ba
 -2-3\,\sqrt {2}+2\,\sqrt {3}-6\, \left( -1/4-1/4\,\sqrt {3} \right)
 \sqrt {2}\\
 1\\
 1/3\,\sqrt {3}\\
 2-2/3\,\sqrt {3}-2\,\sqrt {2}-2\, \left( -1/4-1/4\,\sqrt {3}
 \right) \sqrt {2}
 \ea
 \right )\,.
 \ee
%
%
%
%
%
%
%
Their insertion in the general four-parametric formula (\ref{metrr})
would be straightforward, confirming the efficiency of the
computer-assisted manipulations in dealing with the problems of
quantum catastrophes within the overall PTQM theoretical framework.

\section{Conclusions\label{discussion}}

Via several families of solvable models we showed that besides the
well known role of the numerical calculations and besides the
standard use of graphics, the tools of computer-assisted algebra and
symbolic manipulations might also prove particularly useful and
productive in quantitative and theoretical analyses of many
sufficiently elementary quantum systems and phenomena.

Our main attention has been paid to the study of behavior of a
triplet of toy models near their horizons of observability $\partial
{ \cal D}^{(N)}$. In this light, the physics-oriented readers of
this MS might ask questions about the practical phenomenological
applicability of the models. Let us, therefore, fill the gap and let
us touch this question in this section.

Firstly, let us mention that many years ago, the primary source of
interest in quantum models possessing EP singularities emerged
within the broad framework of perturbation calculations. Although
the most influential Kato's monograph \cite{Kato} on perturbation
series was not strictly aimed at physicists (typically, Barry Simon
bitterly complained, as a student, that he had to study the Kato's
non-Hermitian examples), it left a lasting impact in the field (one
of main reasons was that the EP positions in the complex plane of
couplings determined, mathematically, the radii of convergence of
the series).

Many years later, the majority of quantum physicists still
considered the Kato's introduction of his EP concept via $N=2$ real
matrices rather formal. Nevertheless, these special definitions of
the simplest possible square-root-type EP singularities found in
fact numerous applications in a fairly broad variety of quantum as
well as non-quantum physical contexts (cf., e.g., the webpage
\cite{Heissb} of a recent dedicated international conference).

In comparison, not too much attention has still been paid to the EP
singularities of the higher degeneracy. In many practical
realizations of these degeneracies there emerges a fine-tuning
problem because the mere random variation of phenomenological
parameters offers just very small chances of a fusion of the simple
EP singularities (forming, at $N>2$, the higher-order-EP or multiple
EP {\it alias} MEP degeneracies).

The latter argument looks persuasive \cite{Heiss}. It seems to have
kept the theoretical studies as well as experimental simulations of
the MEP-related phenomena out of the mainstream in contemporary
physics until very recently. Currently, the scepticism may be
weakening. A particularly encouraging physics-oriented sample of the
potentially important applications of the MEP-related ideas has been
presented, for example, in  preliminary report~\cite{preprint}. In
this text one of us argued that the formalism of PTQM might be able
to offer a long-needed innovative conceptual ground for the
unification of the classical Big Bang evolution scenario with the
standard principles of quantum theory in theoretical cosmology.

In paper I \cite{I} the latter idea was prepared and developed
beyond the simplest examples. In order to make any attempted
application feasible one still had to restrict attention to the
Hilbert spaces with finite dimensions. On conceptual level this
already enabled one of us to establish, i.a., one of the first
successful connections of quantum mechanics with the classical
Thom's theory of catastrophes (cf. \cite{Thom}).

The key mathematical message of paper I was that the availability of
the optional and, in principle, variable metric $\Theta^{(S)}$ which
defines the inner product in the  standard Hilbert space ${\cal
H}^{(S)}$ opens the way towards an innovated, deeply
quantum-theoretical concept of a {\em quantum} catastrophe. During
such a catastrophe the very Hilbert space ${\cal H}^{(S)}$ {\em
ceases to exist}. In the present continuation of the theory we
complemented paper I by a number of further models and by the
description of a few technical details including also illustrative
explicit constructions.

In a slightly sceptical mood we should add that in contrast to the
discrete anharmonic oscillator model as used in paper I, the present
three new discrete-lattice models proved slightly less user
friendly. The truly unique and exceptional close-form MEP roots of
paper I had to be replaced, in places, by the mere numerical roots
of the analogous effective polynomials (cf., e.g., the end of
paragraph \ref{firstmo} for our present illustrative example).

In the broader physical area of a realistic phenomenology a lot of
work still remains to be done as well. The main reason is that at
the present stage of development the PTQM approach is still
characterized by the technical difficulties given by the fact that
$\Theta \neq I$. Anyhow, we believe that a systematic analysis of
these difficulties is both necessary and feasible. In this sense,
one of our main conclusions is that for the time being, the best
chances of an imminent progress are still just in a continued study
of the finite-dimensional Hilbert spaces and of the special families
of models.

%
%
%


\section*{Appendix A: A brief account of the PTQM formalism}

\subsection*{A.1. Introduction of sophisticated Hilbert
spaces ${\cal H}^{(S)}$ with nontrivial physical inner products}

In the Bender's and Boettcher's PTQM quantum theory one replaces
Eq.~(\ref{SEti}) by a less usual but, presumably, ``friendlier''
Schr\"{o}dinger equation
 \be
 {\rm i}\partial_t\,|\psi^{(F)}\kt=H\,|\psi^{(F)}\kt\,
  \label{SEtiF}
 \ee
where $|\psi^{(F)}\kt\,  \in \,  {\cal H}^{(F)}\,$ and where the
Hamiltonian is allowed non-Hermitian, $ H \neq H^\dagger$. It is
only required that $H$ possesses a real and non-degenerate
bound-state-type spectrum. The basic mathematical idea of the PTQM
theory may be then formulated as a natural rejection of the
``first'' Hilbert space ${\cal H}^{(F)}$ as ``false'' and
unphysical, i.e., as yielding a manifestly non-unitary evolution of
the system in question via Eq.~(\ref{SEtiF}). Once we want to
interpret Eq.~(\ref{SEtiF}) as fully compatible with any textbook on
quantum mechanics, we simply {\em must} move to an amended,
``second'' Hilbert space ${\cal H}^{(S)}$ \cite{Geyer}.

The first condition of consistency of such a procedure is that the
two Hilbert spaces ${\cal H}^{(F)}$ and ${\cal H}^{(S)}$ only differ
from each other by the definition of the respective inner products.
With no difference between the ket-vectors themselves,
$|\psi^{(S)}\kt \ \equiv \ |\psi^{(F)}\kt\ $, the change only
involves the linear functionals (i.e., the so called bra-vectors).
Thus, one simply replaces the common (a.k.a. ``Dirac's'')
anti-linear Hermitian-conjugation operation
 \be
 {\cal T}^{(F)}: |\psi^{(F)}\kt\ \to \br \psi^{(F)}|
 \label{Dirac}
 \ee
valid in ${\cal H}^{(F)}$ by its alternative
 \be
 {\cal T}^{(S)}: |\psi^{(S)}\kt\ \to \bbr \psi^{(S)}|\ \equiv \
 \br \psi^{(F)}|\,\Theta
 \,
 \ee
valid in the ``standard'' physical Hilbert-space ${\cal H}^{(S)}$
(cf. also \cite{SIGMA} for more details). The purpose of the
introduction of the latter general formula which contains an
optional Hilbert-space metric operator $\Theta=\Theta^\dagger > 0$
is that an appropriate choice of the metric $\Theta$ {\em must} make
our Hamiltonian $H$ self-adjoint in  ${\cal H}^{(S)}$.
Mathematically speaking \cite{recurrent} one must merely guarantee
the validity of the Dieudonn\'e's compatibility condition
 \be
 H^\dagger\,\Theta=\Theta\,H\,.
  \label{dieudo}
 \ee
The key problem of the assignment of a suitable Hermitizing metric
$\Theta$ to a given non-Hermitian Hamiltonian $H$ is that the
Dieudonn\'e's constraint (\ref{dieudo}) still leaves the metric
$\Theta=\Theta(H)$ (and, hence, the physical inner product {\em and}
the physical Hilbert space ${\cal H}^{(S)}$) non-unique.
Fortunately, according to the older review \cite{Geyer} the
resolution of the problem is imminent. One merely specifies a
$K-$plet of additional observables ${\cal C}_j$, $j=1,2,\ldots,K$
with such a count $K$ that the metric (which makes them {\em all}
self-adjoint) becomes unique.

\subsection*{A.2. ${\cal PT}-$symmetric quantum systems}

A hidden reason of the recent success and appeal of the PTQM
formalism in applications \cite{Carl} may be traced back to its
trademark use of an {\em additional}, heuristically enormously
productive assumption
 \be
 H^\dagger {\cal P}={\cal P}\,H\,.
 \label{ptsy}
 \ee
This means that our Hamiltonian operator $H$ is required
self-adjoint in another, associated Krein  space ${\cal K}$ which is
defined as endowed with an indefinite metric (i.e., pseudo-metric)
${\cal P}$ \cite{ali,Langer}. Relation (\ref{ptsy}) re-written in
its formally equivalent and, among physicists, perceivably more
popular representation $H {\cal PT}={\cal PT}\,H\,$ explains the
words ``${\cal PT}-$symmetry of $H$''.

The uniqueness of the physical interpretation of ${\cal
PT}-$symmetric quantum models is often based on the  $K=1$ postulate
of existence of a single  ``missing observable'' ${\cal C}_1$,
exhibiting all of the characteristics of a charge (cf. \cite{Carl}
for a virtually exhaustive account of this version of the theory).
In our present paper a less widespread strategy of making the metric
unique is taken from paper I.

\section*{Appendix B. BIM - GPM correspondence at large $N$}

In Eq.~(\ref{toymSE}) the inner subvector $Q|\psi\kt$ with $(N-4)$
components $\{\psi_3, \ldots, \psi_{N-2}\}$ may be interpreted as an
exact solution of an incomplete set of the standard discrete
free-motion conditions
 \be
 -\triangle \psi_k=E\,\psi_k\,,\ \ \
 k=3, 4, \ldots, N-2\,.
 \label{sfm}
 \ee
Whenever this middle-partition vector $Q|\psi\kt$ is given in
advance, then, on the contrary, the third line of Eq.~(\ref{toymSE})
may be re-read as the definition of component $\psi_2$ while the
$(N-2)$nd line defines $\psi_{N-1}$. At this moment we are left just
with the remaining first two rows of our equation,
  \begin{eqnarray*}
  2\psi_1-(1+\lambda)\psi_2\ \equiv \
  \psi_1+\delta\psi_{3/2}-\lambda\,
  \psi_2 =E\,\psi_1\,,\\
 (-1+\lambda)\psi_1+(2-E)\,
 \psi_2-
 \psi_3\ \equiv\ (-\triangle-E)
\psi_2+\lambda\psi_1= 0\,
  \end{eqnarray*}
plus, {\em mutatis mutandis}, with the last two rows. In terms of
the not yet specified parameter $\delta\psi_{3/2}$ the first row
enables us to define the last ket-vector component $\psi_1$. Once we
eliminate $\psi_1$ from the second row, we obtain
 \be
  (1-E)(-\triangle-E)
\psi_2=\lambda^2\psi_2-\lambda\,\delta\psi_{3/2}\,.
  \ee
At the large dimensions $N \gg 1$, the latter relation may now be
quite naturally re-read as an extension of Eq.~(\ref{sfm}) to $k=2$,
based on a self-consistency condition
 \be
  \lambda\psi_2-\delta\psi_{3/2}=0\,.
  \label{robik}
  \ee
Now one may recall the most natural approximate, large$-N$
identification of the matrix indices $j=1,2,\ldots,N$ with the
suitable equidistant grid points of a real interval. In the
corresponding map $j \to x_j = x_0+jh$, the constant $h>0$ is
assumed small, $h \sim 1/N$.

In the limit of large $N$, the ket-vector elements $\psi_j$ may be
then re-read as the values of a wave function $\psi(x)$ at the
respective coordinate $x=x_j$. Similarly, relation (\ref{robik})
acquires the simplified meaning of the so called Robin's boundary
condition at $x=x_0$,
 \be
 \lambda\psi(x_0)-\partial_x \psi(x_0)= 0\,.
  \label{robikcont}
  \ee
A close connection emerges with the local-interaction model of
Ref.~\cite{david}. It is worth adding that the latter, manifestly
non-Hermitian model found an interesting immediate physical
interpretation via an R-matrix description of one-dimensional
quantum scattering \cite{Coronado,Stefan}. These observations may be
also read as another encouragement of the study of the
multiparametric NNIM generalizations of the one-parametric BIM model
(\ref{toymSE}).

%
%
%


\end{document}